\documentclass[preprint,aps,endfloats,a4paper]{revtex4}

\usepackage{subfigure}
\usepackage{amssymb}
\usepackage{graphicx}
\usepackage{color}
\usepackage{longtable}

\begin{document}

\title{Length dependence of thermal conductivity\\
by approach-to-equilibrium molecular dynamics}
\author{
Hayat ZAOUI, Pier Luca PALLA, Fabrizio CLERI and Evelyne LAMPIN\footnote{evelyne.lampin@univ-lille1.fr}
} \affiliation{
Institut d'Electronique, Micro\'electronique et Nanotechnologies \\(IEMN, UMR CNRS 8520 and University of Lille) \\ Avenue Poincar\'e - CS60069 - 59652 Villeneuve d'Ascq Cedex - France
\\}

\vspace{10 cm}

\begin{abstract}
The length dependence of the thermal conductivity over more than two decades is systematically studied for a range of materials, interatomic potentials and temperatures, by the atomistic approach-to-equilibrium molecular dynamics method (AEMD). By comparing the values of conductivity obtained for a given supercell length and maximum phonon mean-free-path (MFP), we find that such values are strongly correlated, demonstrating that the AEMD calculation with a supercell of finite length, actually probes the thermal conductivity corresponding to a maximum phonon MFP. As a consequence, the less pronounced length dependence usually observed for poorer thermal conductors, such as amorphous silica, is physically justified by their shorter average phonon MFP. Finally, we compare different analytical extrapolations of the conductivity to infinite length, and demonstrate that the 
frequently used Matthiessen rule is not applicable in AEMD. An alternative extrapolation more suitable for transient-time, finite-supercell simulations is derived. This approximation scheme can also be used to classify the quality of different interatomic potential models with respect to their capability of predicting the experimental thermal conductivity. 
\end{abstract}


\maketitle

\section{Introduction}
The thermal properties of materials are modified at the shortest length and time scales, when the characteristic diffusion length becomes comparable with the characteristic system length. Typically, thermal conductivities decrease when approaching the nm-scale\cite{Tang10,Herremans13}, while the electrical conductivity is affected to a much lesser extent, thereby making nanostructured materials good candidates for thermoelectric devices. Moreover, because of the larger surface/volume ratio, the impact of the thermal resistance at the interfaces between different materials becomes concomitantly of increasing importance in nanostructured materials. Both effects combine in integrated circuits reaching the nm-scale, and hamper the dissipation of heat generated in the nm-long channel of the most advanced transistors\cite{Pop10}, thus representing one of the main limits to the further increase of microprocessor operating frequency. Achieving detailed atomistic understanding of thermal transport appears therefore as a crucial prerequisite to overcome such limitations. 

Atomistic simulations can help to understand which physical parameters control the heat path when short length scale effects become important. In particular, molecular dynamics (MD) simulations contain all the ingredients to implicitly simulate the collisions of phonons at the origin of the heat conduction. They can be viewed as computer experiments on a well controlled atomic structure, in a configuration that can be exploited to extract information about thermal properties. In the EMD approach (Equilibrium MD)\cite{Zwanzig65}, the oldest among the MD variants of thermal conductivity calculation, the atomistic system is equilibrated at a given temperature and the thermal conductivity, defined as the same proportionality coefficient between the heat current and the temperature gradient as in the Fourier's law, is deduced from the  fluctuation-dissipation theorem. In the ``direct" method\cite{Schelling02}, on the other hand, a steady-state heat current is established between a heat source and a heat sink, and the linear gradient of the temperature profile is used to extract the thermal conductivity, again according to the Fourier's law. In a different, time-dependent approach, temperature transients can also be used to study the thermal response.\cite{Shenogin04,Hulse05} Recently,\cite{Lampin12,Lampin13} we have shown that when a simulated temperature pulse establishes a step-difference temperature profile in a material, the transient to the equilibrium temperature is exponential, making it easy to extract a typical decay time of the thermal pulse. Using the heat equation, the thermal conductivity can be obtained from this decay time. The advantage of this method, which we called ``approach-to-equilibrium molecular dynamics", or AEMD, is that time transients occur faster, compared to both the attainment of a stationary regime of thermal conduction across a spatial gradient, as in the ``direct" method, and to the numerical convergence required for the Green-Kubo relation in the EMD method. Therefore, much larger system sizes can be studied with AEMD, containing up to 4.5 millions atoms \cite{Lampin13}, and with length $L_Z$ as long as 0.1 mm for a graphene two-dimensional supercell \cite{Barbarino15}. 

Studying large systems is important not only to push the limits of the MD methods, but first and foremost because a pronounced length dependence of the thermal conductivity $\kappa(L_Z)$ is observed in good thermal conductors like silicon. The objective of the present paper is to study the physical origin of this length dependence, and to verify how well the underlying physics is captured by the computer simulations. In the first part of the work, we demonstrate that the length dependence in the time-transient model, embodied by the AEMD, originates from the cut-off on the maximum phonon mean-free-path (MFP), imposed by the finite supercell length. By choosing silicon as the reference material, described by the well-established EDIP interatomic potential,\cite{EDIP} we compare the conductivity corresponding to a given supercell length, $\kappa(L_Z)$, as obtained by AEMD, to that corresponding to a given maximum value ($\Lambda_\mathrm{max}$) of the phonon MFP distribution, $\kappa(\Lambda_\mathrm{max})$, as obtained by another atomistic approach.\cite{Henry08} The strong correlation between the two values allows to explain the evidence of a length dependence observed in the AEMD simulations, in which no other sources of scattering besides the phonon-phonon terms are present. In the second part of the work, we propose an extrapolation of the $\kappa(L_Z)$ curves, based on the above understanding; this is necessary, since the Matthiessen-like relation proposed by Schelling et al. \cite{Schelling02} for the ``direct" method can not be physically justified in the context of AEMD, in which no explicit boundaries could provide a length-dependent phonon scattering. By applying the new formula to the data obtained by different interatomic potentials for crystalline silicon, we highlight the ability of EDIP to quantitatively reproduce the experimental values  of thermal conductivity. Finally, we show that the good agreement of our results with the thermal conductivity values obtained by EMD (a method not subject to a length dependence, once the number of phonon modes, i. e. the number of atoms, is large enough), rules out the existence of 1D effects, which might originate from the extreme aspect ratio of our atomistic structures.

\section{Method}
\begin{figure}[t]
\includegraphics[width=0.75\linewidth]{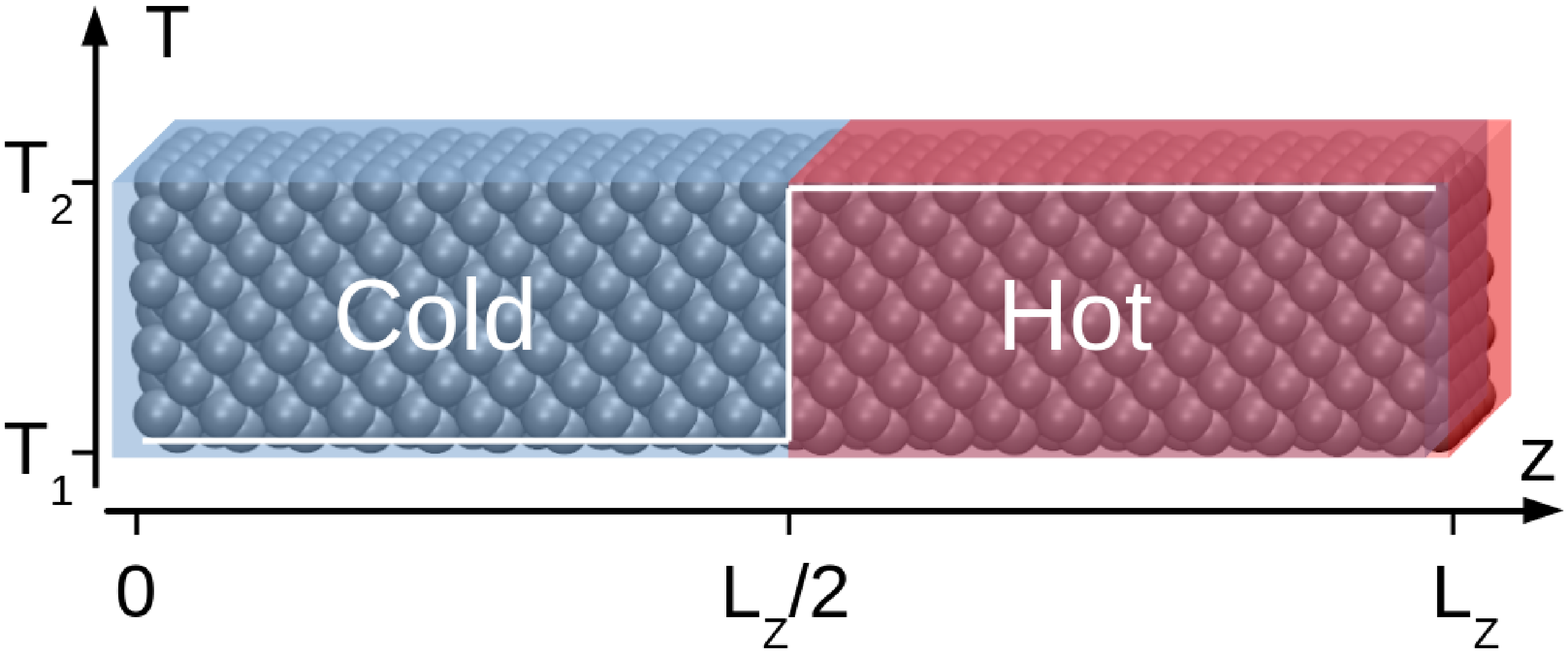}
\caption{Cold and hot blocks in an elongated supercell.}
\label{FigHotCold}
\end{figure}

The principle of the AEMD method is to create and monitor a temperature transient, and use the decay time of the temperature difference to obtain the thermal conductivity of the system. In the present study, devoted to bulk systems, we build atomistic lattices of Si, Ge and SiO$_2$ of finite size $L_X$, $L_Y$ and $L_Z$, initially equilibrated at zero pressure with periodic boundary conditions applied in the three directions. The presence of periodic boundaries represents a periodicity not only in the real space, but also in the dual space of wave vectors, thus giving rise to the concept of ``supercell". The length $L_Z$ is chosen to be much larger than the cross section of the supercell, in order to induce a one-dimensional heat current along $z$. Two Nos\'e-Hoover thermostats\cite{NoseHoover} are used to split the supercell into a cold block at temperature $T_1$ for $0<z<L_Z/2$, and a hot block at temperature $T_2$ for $L_Z/2<z<L_Z$ (Fig. \ref{FigHotCold}). 

\begin{figure}[b]
\includegraphics[width=\linewidth]{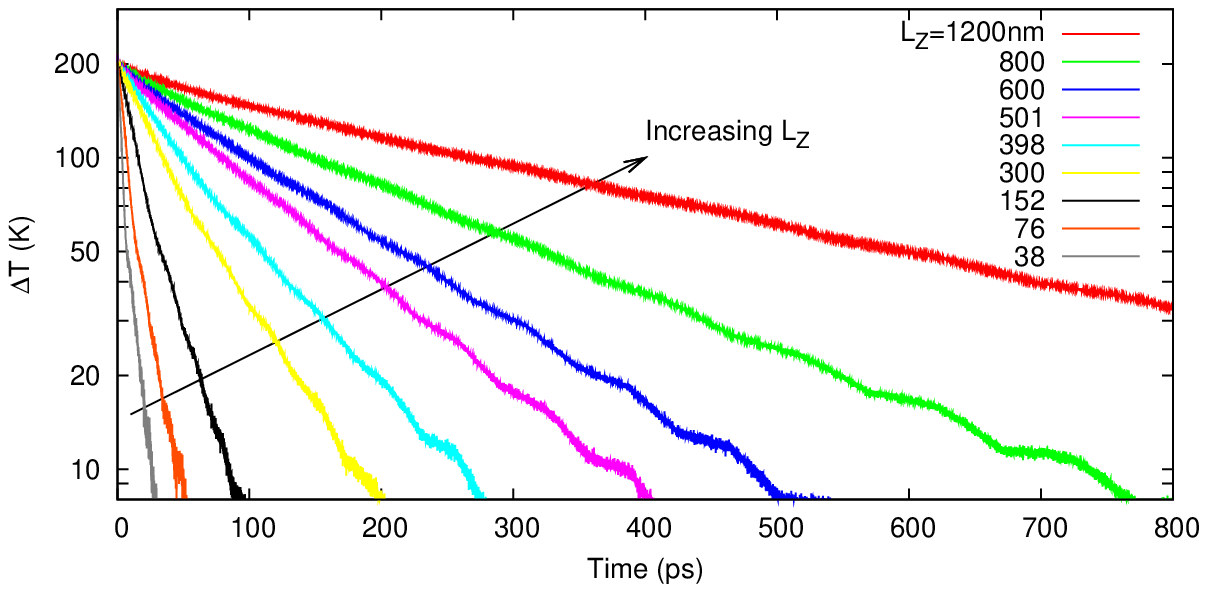}
\caption{Temperature difference between the hot and cold blocks for $L_Z$ ranging from 38 to 1200 nm. Si, Tersoff potential, 300 K.}
\label{FigTransient}
\end{figure}

A significant initial temperature difference $T_2-T_1= 200$ K is chosen, to enhance the amplitude of the decay signal in the subsequent equilibration phase, in order to get a better precision on the transient characteristic time $\tau$. After typically 100 ps of MD simulation at constant-$\{NVT\}$, the two thermostats are removed and the system is left free to reach equilibrium, at constant-$\{NVE\}$. During the approach to equilibrium, the average temperature in each block is monitored, and the difference tends to 0 according to an exponential decay as shown in Fig. \ref{FigTransient}.

This configuration corresponds to the Sommerfeld heat conduction problem on a ring,\cite{Sommerfeld} with a dominant decay constant that is related to the thermal conductivity by the relation:
\begin{equation}
\kappa=\frac{L_ZC_V}{4\pi^2S}\frac{1}{\tau}
\label{EqAEMD}
\end{equation}    
with $C_V$ the heat capacity, determined from separate MD simulations, and $S$ the area of the cross-section of the supercell perpendicular to $z$. 

The length dependence is then studied by varying the supercell length $L_Z$. The same methodology is applied to a range of materials, from good (Si) to poor heat conductors (amorphous SiO$_2$), passing by intermediate materials such as Ge and $\alpha$-quartz. Moreover, the dependence on the interatomic potential is studied for the case of Si and three interatomic potentials, frequently used to describe the Si-Si interactions: the Tersoff, \cite{Tersoff88} the Stillinger-Weber,\cite{SW85} both with the original parametrisation and the Lee and Hwang's parameters\cite{Lee12}, and the EDIP. The Ge-Ge interactions are modelled by the Tersoff potential. In $\alpha$-quartz, the interactions are modelled by the BKS potential. \cite{BKS} Amorphous silica is modelled using Munetoh's parametrisation\cite{Munetoh07} of the Tersoff potential for both the Si-O and O-O interactions, while Si-Si interactions are set to zero; the amorphous structure is obtained from the quench of molten silica.\cite{Lampin12} The impact of the temperature is studied in the case of Si described with Tersoff and EDIP potentials. 
  
\section{Length dependence of thermal conductivity}

\subsection{Supercell size vs. phonon mean free path in Si}

\begin{figure}[tb]
\includegraphics[width=\linewidth]{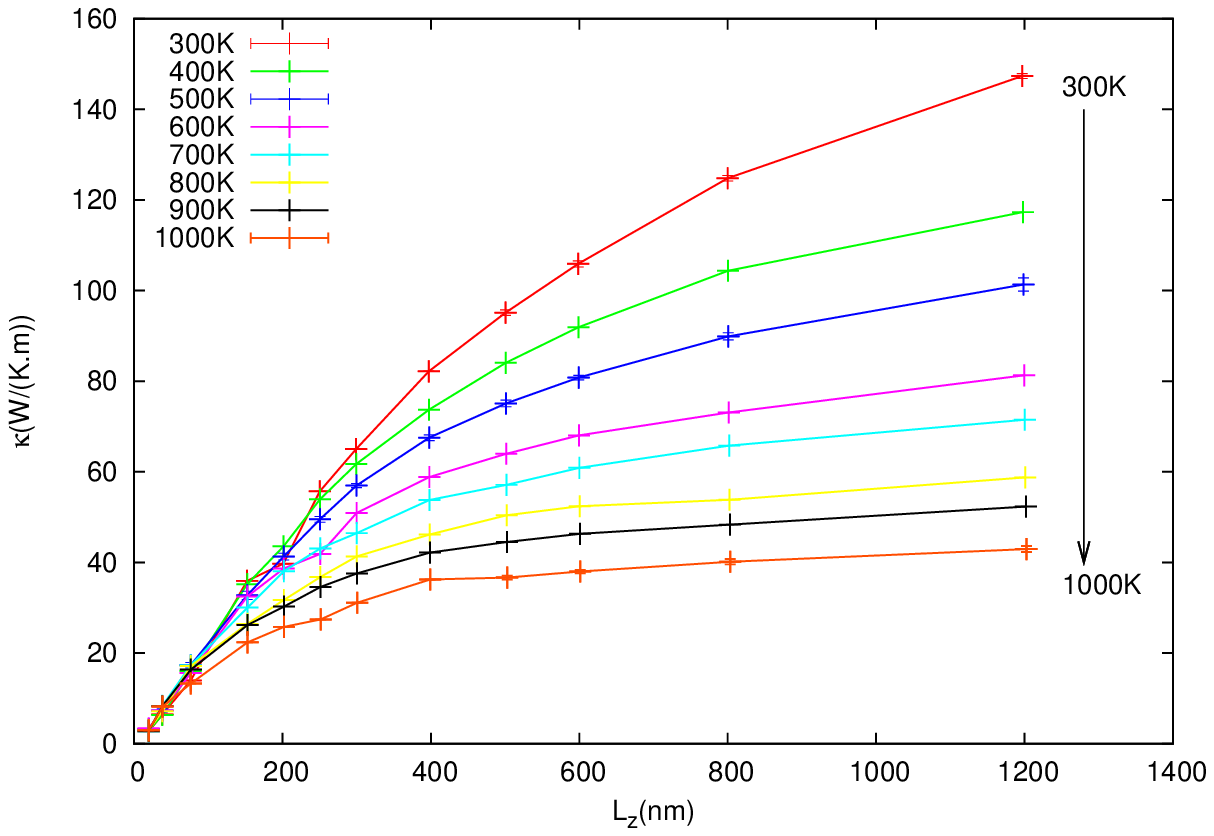}
\caption{Thermal conductivity $\kappa$ of silicon versus length $L_Z$ for temperatures ranging from 300 to 1000 K. Tersoff potential.}
\label{FigConTersxLogwof}
\end{figure}

The length dependence of the thermal conductivity $\kappa(L_Z)$ is firstly studied for the already well-documented case of bulk crystalline silicon, described by the Tersoff interatomic potential. In this case, the supercells have a section of 16$\times$16 lattice units, corresponding to 8.7$\times$8.7 nm$^2$ at 300 K. The length is varied from 70 to 2200 lattice units, corresponding to 38 to 1200 nm at 300 K. The decay time $\tau$ increases with the length, as can be seen from the curves in Fig. \ref{FigTransient} at 300 K. A fine determination of $\tau$ and of the error is obtained by averaging over many exponential fits of each decay curve, from an initial time $t_0\in[0,t_e/2]$, to $t_e$ the end of the simulation. By this procedure, the initial fast transient corresponding to the switching from $\{NVT\}$ to $\{NVE\}$ (actually an artefact of the MD simulation) can be removed, and the intrinsic thermal decay time is recovered.
The decay time is used to calculate the thermal conductivity from Eq. \ref{EqAEMD} and the resulting values are presented in Fig. \ref{FigConTersxLogwof}.
 The target temperature, representing the averages of the hot and cold initial values, covers the interval from 300 to 1000 K, in steps of 100 K. For the smallest sizes ($L_Z\leq150$ nm), the value is obtained by averaging 7 simulations initialised with different atomic velocities following a Maxwell-Boltzmann distribution at that temperature. Moreover, at T = 300, 500 and 1000 K, the statistical analysis is extended to the whole range of lengths shown in Fig. \ref{FigConTersxLogwof}. Notwithstanding this accurate statistical analysis, the corresponding error bars are small and are not visible on the graph. 

The calculated thermal conductivity strongly depends on the supercell length over the whole temperature range. The saturation to a constant, length-independent value of $\kappa$ cannot be obtained, even for the longest cell ($L_Z$ = 1.2 $\mu$m). In addition, the saturation value is closer to the maximum simulation length that we could study, upon increasing the temperature: from $L_Z=0.8$ to $1.2$ $\mu$m, the thermal conductivity increases by 15\% at 300 K, 10\% at 500 K and 7\% at 1000 K. Meanwhile, the phonon MFPs, $\Lambda$, decrease upon increasing temperature, because the atomic vibrations begin to sample regions of the energy landscape well away from the harmonic minimum. In order to find if these two effects are related, we compared our curves $\kappa(L_Z)$ to the $\kappa(\Lambda_\mathrm{max})$ curves of thermal conductivity as a function of the maximum phonon MFP by Henry and Chen.\cite{Henry08} In that work, they calculated the thermal conductivity from the Boltzmann equation, by integrating over the phonon wavevector \textbf{k}, and summing over the phonon modes $\nu$:
\begin{equation}
\kappa_{\mathrm{bulk}}=\frac{1}{V}\sum_{\nu}\int\hbar \omega_{\nu}v_{\nu}(k)\Lambda_{\nu}(k)\frac{\partial f_{\mathrm{BE}}(\omega_{\nu},T)}{\partial T} d\mathbf{k}
\label{EqBoltz}
\end{equation}
where $\hbar \omega_{\nu}$ is the energy of the mode $\nu$, $v_{\nu}(k)$ is the group velocity, $f_{\mathrm{BE}}$ is the Bose-Einstein distribution of phonons at the temperature $T$, and $\Lambda_{k,\nu}=\tau_{k,\nu}v_{k,\nu}$ is the mean free path, with $\tau_{k,\nu}$ the relaxation time of the mode. Henry and Chen carried the integration over the phonon frequencies by using the density of states $D(\nu)$ as input. Lattice dynamics calculations were carried out to obtain the phonon frequencies and eigenvectors from the dynamical matrix, the energy-dependent group velocities, and the density of states. The relaxation times on the other hand, were obtained from MD trajectories transformed to normal-mode coordinates (see Ref. \onlinecite{Henry08} for more details).  The calculation of the thermal conductivity was performed along the directions [100], [110] and [111] and finally averaged to get the bulk thermal conductivity $\kappa_{\mathrm{bulk}}$. They also performed partial integrations from 0 to $\nu=\nu_\mathrm{max}$ to obtain the accumulation curve $\kappa(\Lambda_\mathrm{max})$, $\Lambda_\mathrm{max}$ being the average of $\Lambda(\nu_\mathrm{max})$.

\begin{figure}[htb]
\includegraphics[width=\linewidth]{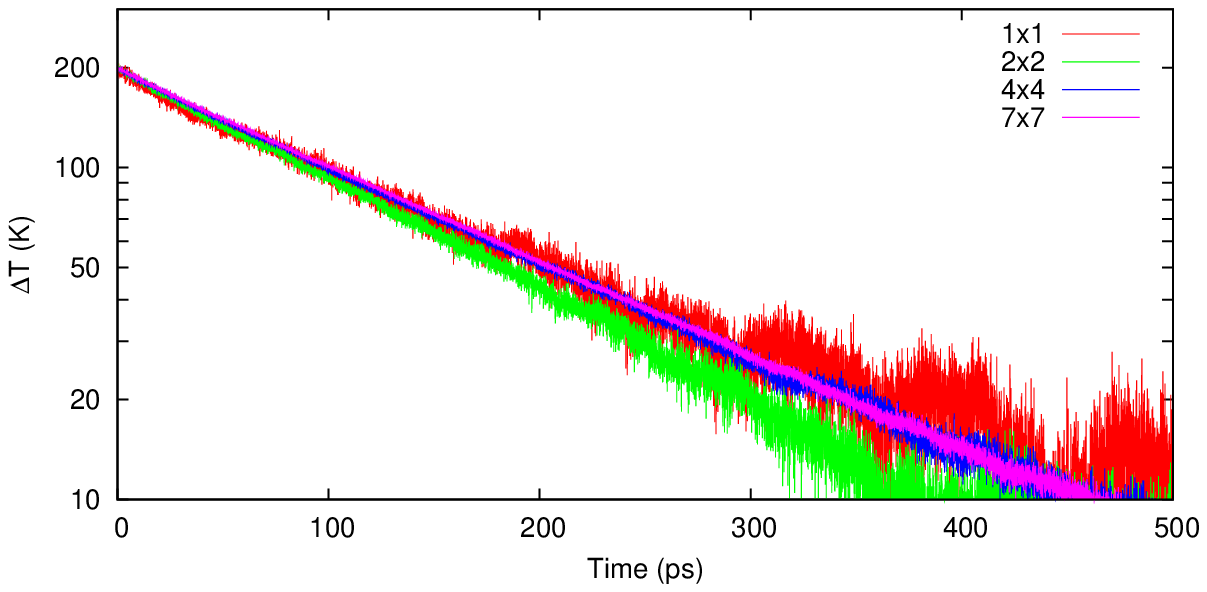}
\caption{Effect of the cross section (in lattice units) on the temperature transient for a length of 520 nm. EDIP potential, 300 K.}
\label{FigSection}
\end{figure}

The calculations by Henry and Chen were performed with the EDIP potential, and they obtained bulk thermal conductivities of 175 and 25 W/(K.m), respectively at 300 and 1000 K. We performed calculations by AEMD using the same interatomic potential at 300, 500 and 1000 K. The cross section of our supercells is equal to 7$\times$7 lattice units except for the largest supercells ($L_Z>1$ $\mu$m) that were only achievable using a smaller section of 4$\times$4 lattice units. We checked at smaller $L_Z$ (520 nm) that the decay transients only differ by less than 1\% between the 7$\times$7 and 4$\times$4 case, and that the cross section could not be further decreased without significantly increasing the noise (Fig. \ref{FigSection}).

\begin{figure}[htb]
\includegraphics[width=\linewidth]{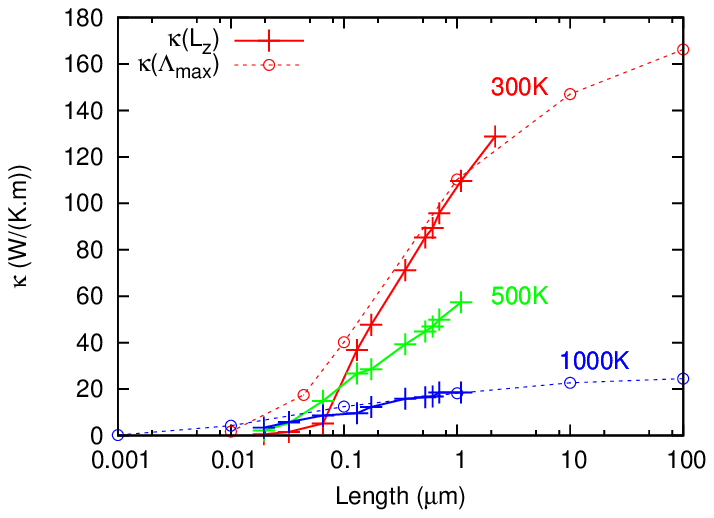}
\caption{Thermal conductivity by AEMD versus supercell length $\kappa(L_Z)$ and from Ref. \onlinecite{Henry08} versus maximum MFP $\kappa(\Lambda_\mathrm{max})$. Temperatures of 300 (red), 500 (green) and 1000 K(blue).}
\label{FigConEdip}
\end{figure}

\begin{figure}[htb]
\includegraphics[width=\linewidth]{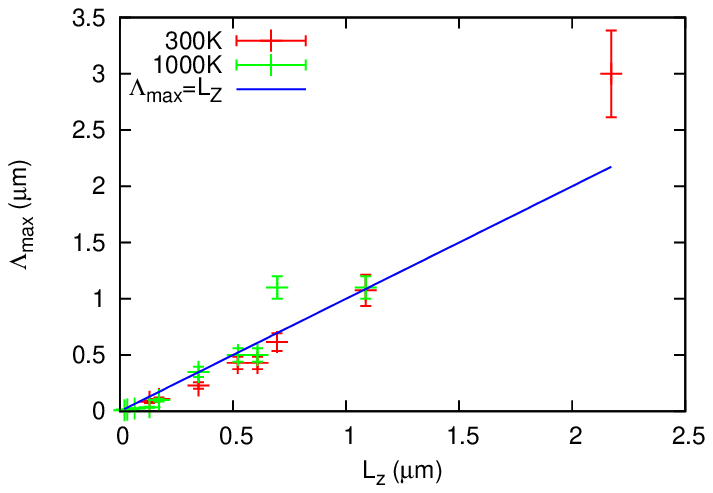}
\caption{Maximum phonon MFP versus supercell length. EDIP, 300 and 1000K.}
\label{FigMFPvsLz}
\end{figure}

The thermal conductivity $\kappa(L_Z)$ obtained by AEMD is plotted in Fig. \ref{FigConEdip}  together with the MFP-dependent conductivity by Henry and Chen $\kappa(\Lambda_\mathrm{max})$.
Fig. \ref{FigConEdip} shows a very similar quantitative behavior of $\kappa(L_Z)$ and $\kappa(\Lambda_\mathrm{max})$. In order to better quantify the relation between the supercell length and the phonon MFPs, we have determined for each value of $L_Z$, the $\Lambda_\mathrm{max}$ corresponding to the same value of thermal conductivity. The result is plotted in Fig. \ref{FigMFPvsLz}.
A strong correlation is obtained, thus demonstrating that the $\kappa(L_Z)$ curves obtained by AEMD indeed probe the phonon MFP distribution, through an accumulation function whose upper-MFP cut-off coincides with the value of $L_Z$. This also gives support to other works \cite{Barbarino15}, also based on our AEMD method, in which such a relationship was taken for granted, but not fully demonstrated.

\subsection{Comparison between ``good" and ``bad" conductors}

\begin{figure}[htb]
\includegraphics[width=\linewidth]{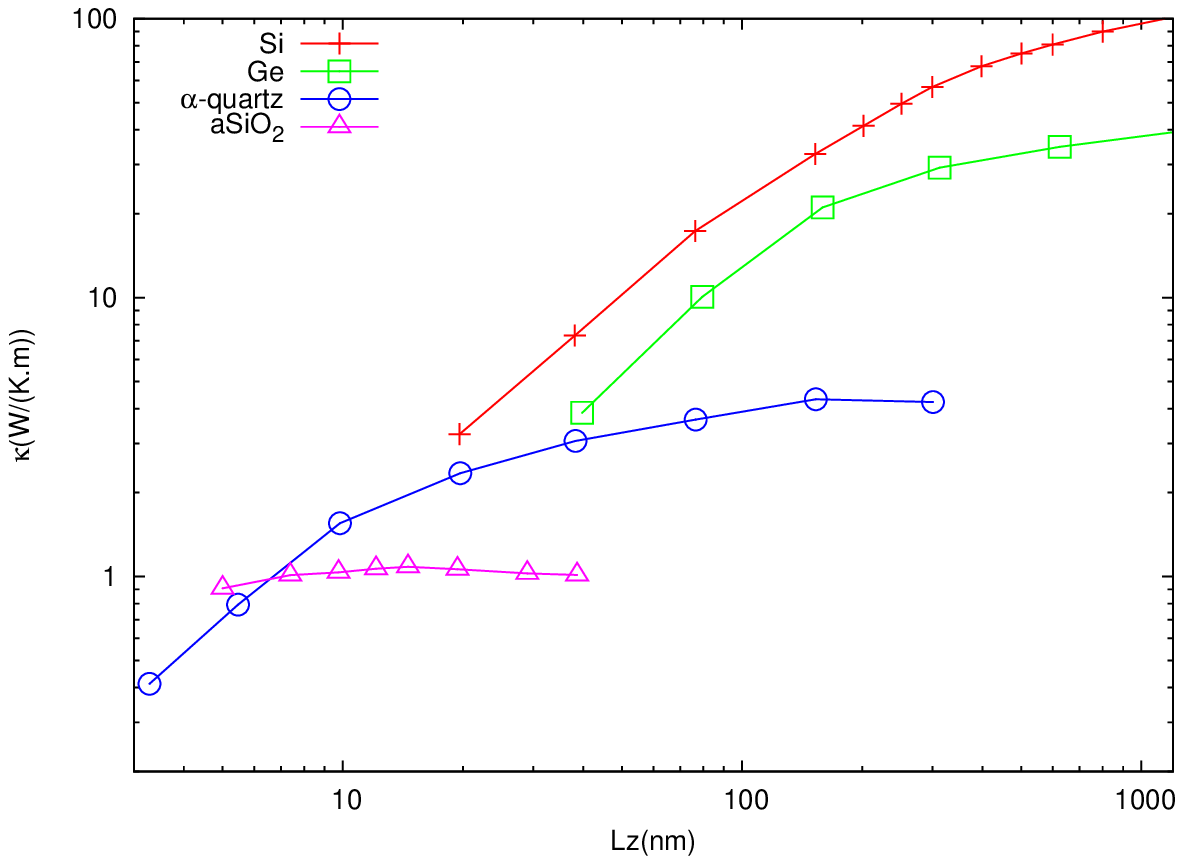}
\caption{Thermal conductivity versus length for silicon, germanium, $\alpha$-quartz, and a-SiO$_2$. 500 K.}
\label{FigCondLogLog500}
\end{figure}

We have studied the $\kappa(L_Z)$ dependence in various materials (Fig. \ref{FigCondLogLog500}).
 The target temperature is equal to 500 K in the four cases. The supercell cross section is equal to 16$\times$16 lattice units for Si and Ge, 18$\times$18 hexagonal units for $\alpha$-quartz and 4.7$\times$4.7 nm$^2$ for a-SiO$_2$. 

The infinite-length thermal conductivity can be obtained for silica. In a-SiO$_2$ the notion of phonon can not be strictly used due to the atomic disorder, and heat carrier MFPs are expected to be very short. The pink curve in Fig. \ref{FigCondLogLog500} shows a plateau at $\approx$ 1 W/(K/m) on the whole length range, which indicates maximum MFP in the nm-range. The thermal conductivity of crystalline $\alpha$-quartz changes by less than 3\% for $L_Z>150$ nm, which indicates the existence of phonon MFPs larger by two order of magnitude compared to a-SiO$_2$. Larger maximum MFPs are indeed expected in this crystalline phase of silica, where the scattering is not limited by the atomic disorder. The infinite-length thermal conductivity is not attained in crystalline germanium, suggesting the existence of $\mu$m-long phonon MFPs. The coupled behavior of maximum MFP and thermal conductivity is  perfectly consistent with the deductions from the kinetic theory: the thermal conduction is poorer in materials where phonon scattering is higher or, in other words, where phonon MFPs are smaller.

In conclusion, the more pronounced length dependence obtained for the lower temperatures, and/or for increasingly better thermal conductors, can be explained by the larger phonon MFPs distribution. In the AEMD method, the supercell length acts as a cut-off in the maximum MFP accounted for in the simulations, and this provides a way to probe the phonon MFPs distribution. Notably, the AEMD method can be applied to the whole range of conductors, despite the longer transients in the case of poor conductors, since the decay time is inversely proportional to the thermal conductivity (Eq. \ref{EqAEMD}). The latter drawback is compensated by a less pronounced, or even nonexistent length dependence of $\kappa$, which makes it possible to determine the thermal conductivity with a computational cost comparable to that required for better thermal conductors such as Si.

\section{Extrapolating the bulk thermal conductivity}

We have shown in the previous Section that the length dependence of the thermal conductivity in AEMD can be  interpreted as a cut-off on the phonon MFP distribution. In this Section a physical interpretation of this dependence is presented.

Notably, a clear length-dependence of the thermal conductivity is always observed when boundaries are explicitly introduced, such as a physical materials interface, or the heat source and sink used in the ``direct" MD method. In such cases, the following formulation is used to extrapolate the bulk thermal conductivity:
\begin{equation}
\frac{1}{\kappa_1(L_Z)}=\frac{1}{\kappa_{\mathrm{bulk}}}\left(1+\frac{\lambda}{L_Z}\right)
\label{ExtraSchell}
\end{equation}
where $\lambda$ has the dimension of a length. This formulation, initially proposed by Schelling et al. \cite{Schelling02}, is based on the kinetic theory formulation of the thermal conductivity. The relaxation time is written by means of a Matthiessen rule, combining the phonon-phonon scattering time and a boundary scattering time. This latter term ``naturally" introduces a  $L_Z$-dependence in the thermal conductivity. 

On the other hand, the length dependence of the conductivity could also be given a purely numerical interpretation, as a Taylor expansion of the intrinsically length-dependent quantity $\kappa$, converging at its asymptotic value for $L_Z\rightarrow \infty$. Such an expansion was proposed by Sellan,\cite{Sellan10} under the assumption of computing a distribution of frequency-dependent relaxation times for each phonon mode, to be fitted by a series expansion. The first-order Taylor expansion would coincide with the same formulation in Eq. \ref{ExtraSchell}, and the eventual extension to second-order could be used, to further improve the numerical fit of the length-dependent data:
\begin{equation}
\frac{1}{\kappa_2(L_Z)}=\frac{1}{\kappa_{\mathrm{bulk}}}\left(1+\frac{\lambda}{L_Z}+\frac{1}{2}\left(\frac{\mu}{L_Z}\right)^2\right)
\label{ExtraSchell2}
\end{equation}
with $\mu$ another parameter having as well the dimension of a length. 

\begin{figure}[htb]
\includegraphics[width=\linewidth]{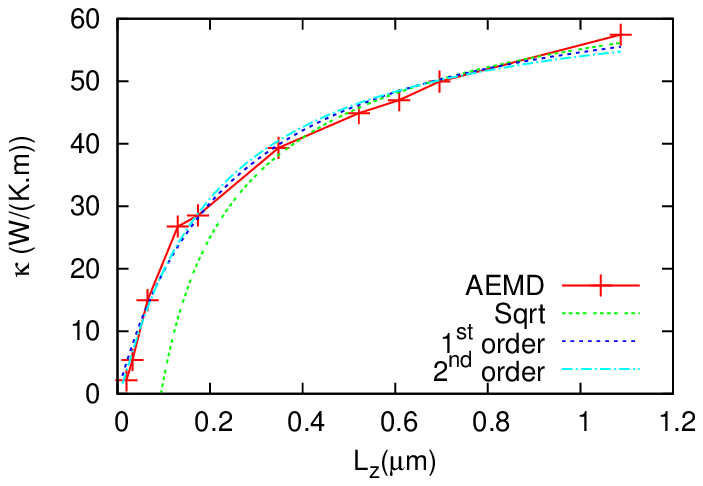}
\caption{Thermal conductivity of Si versus length obtained by AEMD (points), and fit by Eq. \ref{EqFit} (``Sqrt''), by Eq. \ref{ExtraSchell} (``1$^{\mathrm{st}}$ order'') and by Eq. \ref{ExtraSchell2} (``2$^{\mathrm{nd}}$ order''). EDIP potential, 500 K.}
\label{FigTersFit}
\end{figure}

Both formulations have been applied in previous works, to extrapolate at infinite length the thermal conductivity values obtained by AEMD.\cite{Lampin13,Melis14,Dettori15} For the simulations of the present study, using the two fitting procedures gives results as shown in Fig. \ref{FigTersFit}.
Both formulations appear to capture quite well the length dependence, the second-order expansion providing even a slight improvement at short $L_Z$. 

However, in the AEMD approach, both Eq. \ref{ExtraSchell} and \ref{ExtraSchell2} can only be viewed as convenient mathematical fitting functions, since in this method there are no sources of boundary scattering (the supercell is fully periodic), and the apparent length dependence cannot be physically justified by such expressions. The purpose of the present section is to propose a formulation based on a different physical interpretation, compatible with the length dependence obtained in AEMD. 

Starting from Eq. \ref{EqBoltz}, we firstly assume an effective linear relation dispersion, valid at low frequencies ($\omega<\omega_0$), for which $\omega_{\nu} = v_{\mathrm{g}}k$ and $f_{\mathrm{BE}}\approx kT/(\hbar\omega)$. This assumption relies on the demonstration\cite{Broido07,Sellan10b} that the major contribution ($>$ 95 \%) of the thermal conductivity of bulk silicon comes from acoustic phonons. Using this assumption, we now rewrite the integral as a function of the frequency $\omega$:
\begin{equation}
\kappa_{\mathrm{bulk}}=\frac{1}{V}\sum_{\nu}\int_0^{\omega_0}\hbar \omega v_{\mathrm{g}}\Lambda_{\nu}(\omega)\frac{\partial f_{\mathrm{BE}}(\omega,T)}{\partial T} 4\pi \omega^2 d\omega 
\end{equation}
\begin{equation}
\propto \sum_{\nu}\int_0^{\omega_0} \omega^2 \Lambda_{\nu}(\omega) d\omega
\label{Eqkbw}
\end{equation}
The frequency dependence of $\Lambda$ can be taken as  $\Lambda(\omega)\propto\omega^{-n}$, and plugged in Eq. \ref{Eqkbw} to obtain:
\begin{equation}
\kappa_{\mathrm{bulk}}\propto\int_0^{\omega_0} \omega^{2-n} d\omega
\end{equation}
Now we can change again the integration variable to $\Lambda$, obtaining:
\begin{equation}
\kappa_{\mathrm{bulk}}\propto\int_{\Lambda_0}^{\infty} \Lambda^{-3/n} d\Lambda\propto(\Lambda_0^{1-3/n}-\Lambda_{\mathrm{max}}^{1-3/n})
\label{Eqkbulk}
\end{equation}
with $\Lambda_0$ an adjustable parameter, corresponding to the lower limit in the MFP domain where the above assumption of a linear relation dispersion is valid, i. e. $\Lambda_0=\Lambda(\omega_0)$.

We have shown in the previous Section that AEMD simulations at finite $L_Z$ have the consequence of cutting the maximum MFP to a value $\Lambda_\mathrm{max} \simeq L_Z<\Lambda_{\mathrm{bulk}}$. Therefore, the AEMD thermal conductivity at a length $L_Z$ can be interpreted in the same way as Eq. \ref{Eqkbulk}, but terminating the integral at a finite maximum MFP, approximately equal to $L_Z$:
\begin{equation}
\kappa(L_Z)\propto\int_{\Lambda_0}^{L_Z} \Lambda^{-3/n} d\Lambda\propto(\Lambda_0^{1-3/n}-L_Z^{1-3/n})
\end{equation}
 
In this way, the thermal conductivity dependence on the supercell length $L_Z$ (or, equivalently, $\kappa(\Lambda_\mathrm{max})$) reads:
\begin{equation}
\kappa(L_Z) = \kappa_{\mathrm{bulk}}\left(1-\left(\frac{L_Z}{\Lambda_0}\right)^{1-3/n}\right)
\end{equation}
provided $0<n<3$. For phonon-phonon scattering by Umklapp mechanism, the frequency dependence of the MFP is in $\omega^{-2}$, and the interpolating function becomes:
\begin{equation}
\kappa(L_Z) = \kappa_{\mathrm{bulk}}\left(1-\sqrt{\frac{\Lambda_0}{L_Z}}\right)
\label{EqFit}
\end{equation}
Eq. \ref{EqFit} is one particular case of the general forms obtained by Yang and Dames \cite{Yang13} for the accumulation functions.

We have applied this square-root (\emph{Sqrt}) interpolation function to the $\kappa(L_Z)$ values obtained by AEMD simulations. The values of $\kappa_\mathrm{bulk}$ and $\Lambda_0$ for each simulation are given in Tables \ref{TabFitk} and \ref{TabFitL}.
\begin{table}[b]
\caption{$\kappa_\mathrm{bulk}$ (W/(K.m)) obtained by fitting our AEMD results by Eq. \ref{EqFit}.}
\label{TabFitk}
\begin{tabular}{ccccccccc}
Temperature (K) & 300 & 400 & 500 & 600 & 700 & 800& 900 & 1000\\
\hline
EDIP & 169 &-&81 & -&  -& -&  -& 23 \\
Tersoff & 233  & 176&145  & 111&  96&  76&  66& 54 \\
Stillinger-Weber& - &  -&152 & -&  -&  -&  -&- \\
Lee & - &  -&124 & -&  -&  -& -&- \\
\hline
\end{tabular}
\end{table}
\begin{table}[b]
\caption{$\Lambda_0$ (nm) obtained by fitting our AEMD results by Eq. \ref{EqFit}.}
\label{TabFitL}
\begin{tabular}{ccccccccc}
Temperature (K) & 300 & 400 & 500 & 600 & 700 & 800& 900 & 1000\\
\hline
EDIP &  130 & - & 90 & -&  -&  -&  - & 30\\
Tersoff &  188 &137 & 125 &90&74&68&60 & 54\\
Stillinger-Weber&  - & - & 119 & -&  -&  -&  -&  -\\
Lee &  - & -& 98 & -&  -&  -&  -&  -\\
\hline
\end{tabular}
\end{table}
The interpolation using the \emph{Sqrt} formulation is plotted in green in Fig. \ref{FigTersFit} in the case of EDIP calculations at 500 K. The \emph{Sqrt} formulation matches well the results of the AEMD simulations at large $L_Z$, i. e. in the long-wavelength domain for which the above derivation is valid ($L_Z > \Lambda_0$). The values of $\Lambda_0$ (Table \ref{TabFitL}) correspond to the range where significant contributions to the thermal conductivity start accumulating (Fig. \ref{FigConEdip}). The values of the bulk thermal conductivity extrapolated using the the \emph{Sqrt} formulation and the 1$^{\mathrm{st}}$ and 2$^{\mathrm{nd}}$ order formulations are close, although the value is consistently lower with the Eq. \ref{ExtraSchell}. These differences will be discussed in the following. 

\begin{figure}[h]
\includegraphics[width=\linewidth]{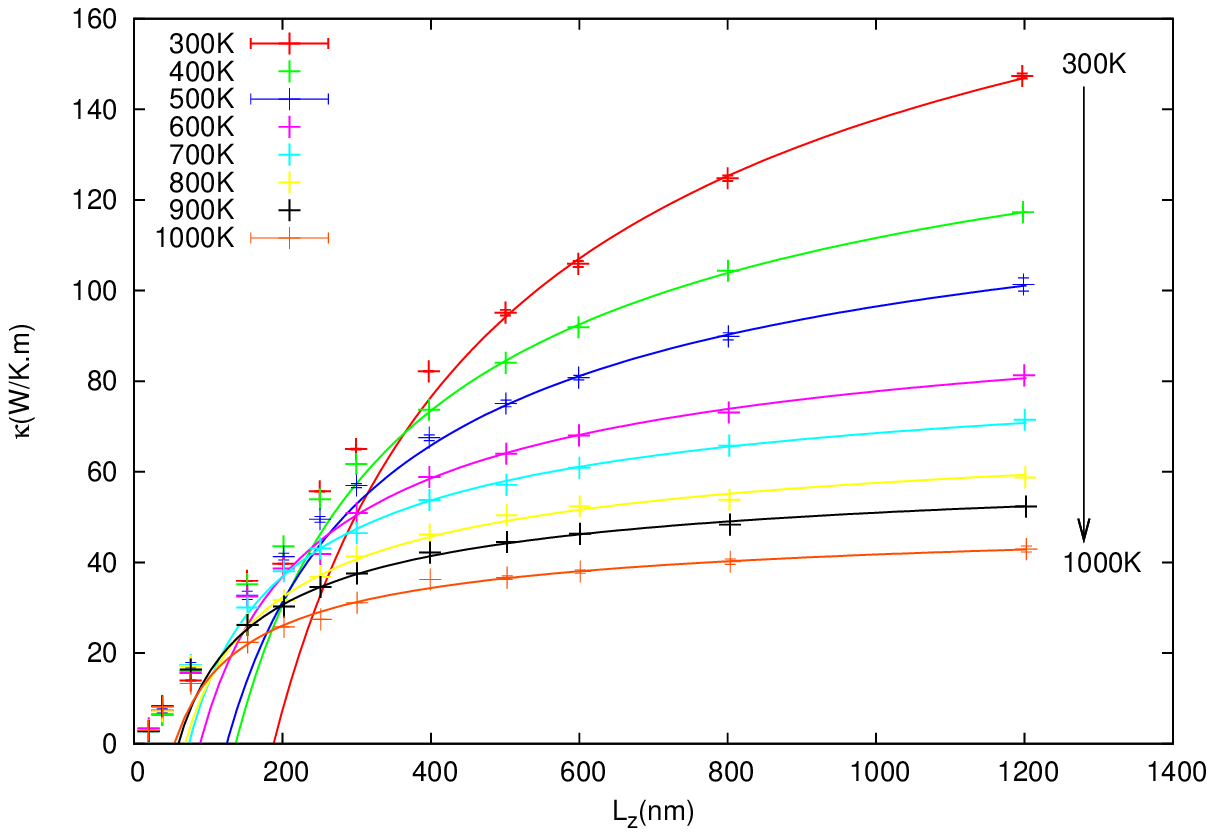}
\caption{Thermal conductivity of Si versus length obtained by AEMD (points) and fit by Eq. \ref{EqFit} (curves). Tersoff potential.}
\label{FitTersoff}
\end{figure}
\begin{figure}[h]
\includegraphics[width=\linewidth]{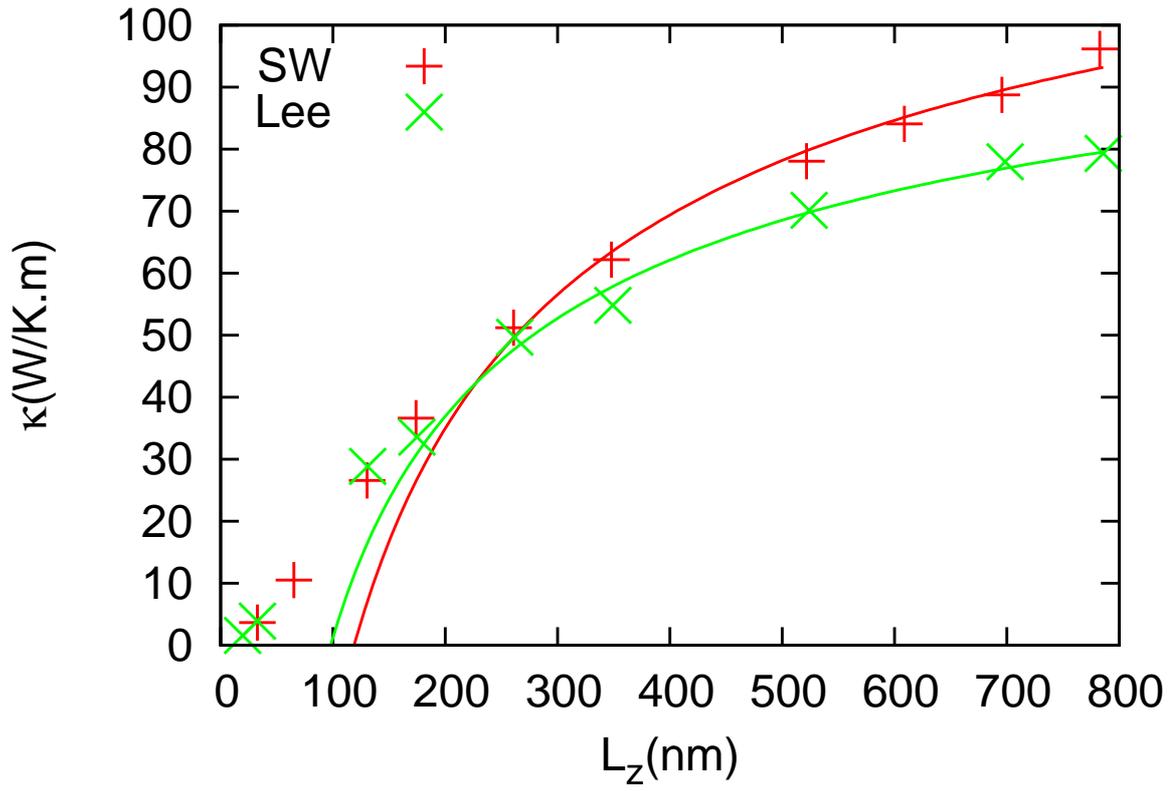}
\caption{Thermal conductivity of Si versus length obtained by AEMD (points) and fit by Eq. \ref{EqFit} (curves). Stillinger-Weber potential, with original (``SW'') and Lee et al.'s parameters (``Lee''), 500 K.}
\label{FitSW}
\end{figure}
\begin{figure}[h]
\includegraphics[width=\linewidth]{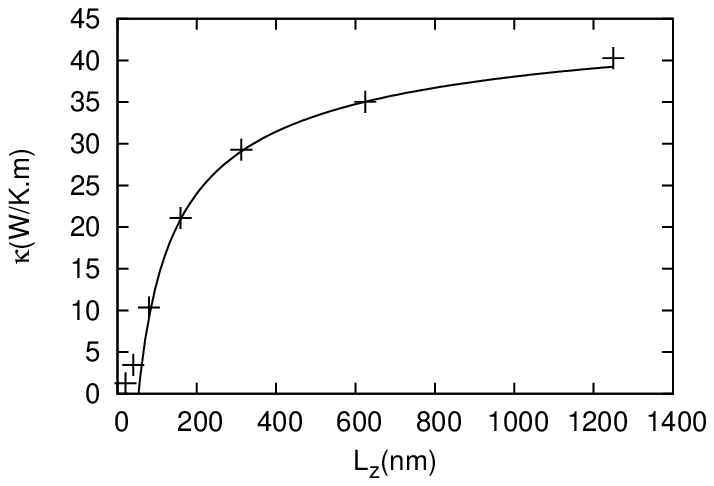}
\caption{Thermal conductivity of Ge versus length obtained by AEMD (points) and fit by Eq. \ref{EqFit} (curves). Tersoff potential, 500 K.}
\label{FitGe}
\end{figure}
\begin{figure}[h]
\includegraphics[width=\linewidth]{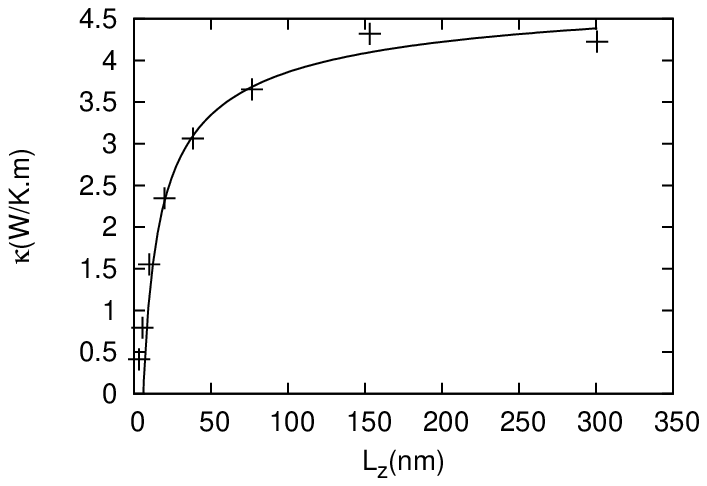}
\caption{Thermal conductivity of $\alpha$-quartz versus length obtained by AEMD (points) and fit by Eq. \ref{EqFit} (curves). BKS potential, 500 K.}
\label{Fitaqu}
\end{figure}

The \emph{Sqrt} interpolation function (Eq. \ref{EqFit}) has also been applied to the AEMD simulations with the Tersoff (Fig. \ref{FitTersoff}) and Stillinger-Weber interatomic potentials (Fig. \ref{FitSW}), as well as for Ge (Fig. \ref{FitGe}) and $\alpha$-quartz (Fig. \ref{Fitaqu}).
The length dependence at large $L_Z$ is again in good agreement with Eq. \ref{EqFit} in all cases: the \emph{Sqrt} model, which relies on two assumptions, namely a length dependence equal to a MFP accumulation, and a phonon MFP due to intrinsic (Umklapp) scattering, appears to capture well the length dependence of AEMD simulations, for a large range of materials and temperatures. 
The fact that the formula fails at very short lengths should not be surprising: the purpose of any fitting function in this context is to provide the best possible extrapolation to infinite length, not to reproduce the MD results. In this sense, the  \emph{Sqrt} formulation has the evident advantage of being physically grounded, instead of being just a numerical fit.\color{black}

\section{Discussion}

A thorough comparison of the thermal conductivity calculated with different methods, different interatomic potentials, and for different materials, has been rarely discussed in the literature\cite{AbsdaCruz11,He12}. This is the purpose of the present Section.

\begin{figure}[htb]
\includegraphics[width=\linewidth]{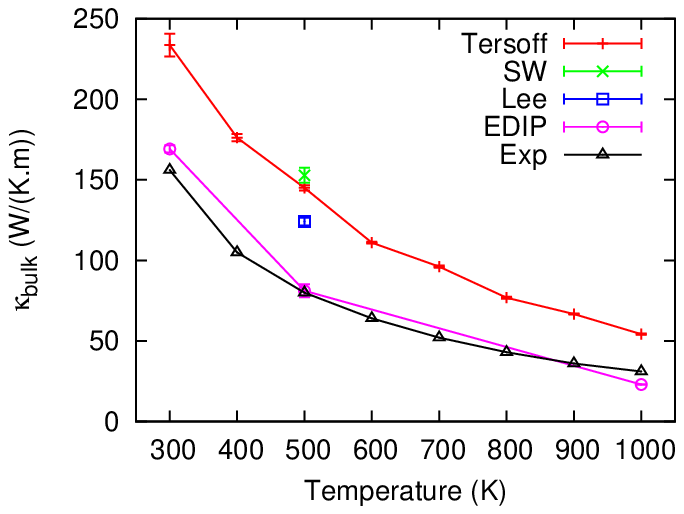}
\caption{Silicon bulk thermal conductivity versus temperature. Calculations by AEMD using various interatomic potentials and experiments \cite{CondSiexp}.}
\label{Figkinf}
\end{figure}

The AEMD bulk thermal conductivity obtained for silicon with the four different interatomic potentials is presented in Fig. \ref{Figkinf}.
It can be seen that a truly quantitative agreement with the measurements can be obtained by using EDIP. The thermal conductivity is overestimated using Tersoff and Stillinger-Weber and, although the parametrisation of the latter potential by Lee and Hwang has some effect on the thermal conductivity, the bond strengthening  proposed in that formulation is not enough to match the experimental conductivity values.

\begin{figure}[b]
\includegraphics[width=\linewidth]{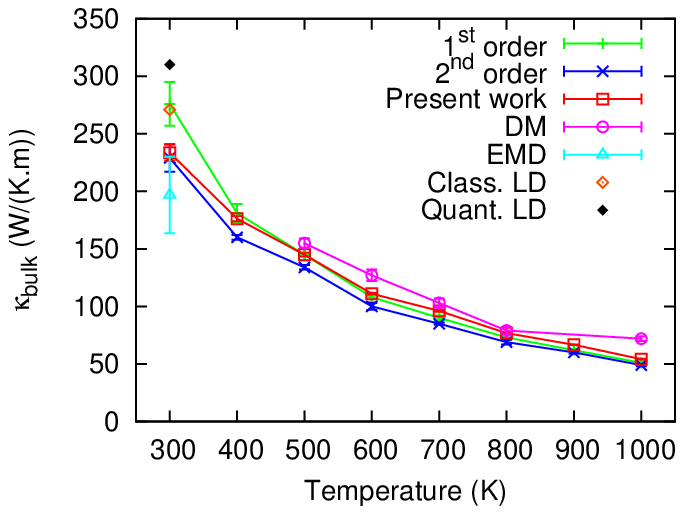}
\caption{Silicon bulk thermal conductivity calculated using Tersoff  potential and various atomistic approaches: AEMD extrapolated with the 1$^{\mathrm{st}}$ and 2$^{\mathrm{nd}}$ order formulations, and by the \emph{Sqrt}. ``Direct method"\cite{Howell12} extrapolated using Eq. \ref{ExtraSchell} (``DM''). EMD, classical (``Class. LD'') and quantum (``Quant. LD'') lattice dynamics calculations \cite{He12}.} 
\label{FigkinfT3S}
\end{figure}

The values of thermal conductivity obtained from different methods, and fitted according to various formulations, can be compared in Fig. \ref{FigkinfT3S}, for the case of Si with the Tersoff potential. Fitting the AEMD $\kappa(L_Z)$ curves by Eq. \ref{EqFit} does not change drastically the value of $\kappa_\mathrm{bulk}$, compared to the value obtained using Eq. \ref{ExtraSchell} and \ref{ExtraSchell2}.  However, it is worth underscoring once more that the \emph{Sqrt} (although valid only at large enough $L_Z$) provides a physically grounded interpretation of the length dependence for these cases (like AEMD) in which no sources of boundary scattering exist.  Also the thermal conductivity obtained by Howell \cite{Howell12}, using the direct method and adjusting the length dependence at first order (Eq. \ref{EqFit}), is comparable to the present results (the residual small difference could be due to a better precision in the present simulations, thanks to larger $L_Z$ used).

Even more interesting is the comparison with the results from methods that do not present an intrinsic length dependence, such as EMD and lattice dynamics. The results obtained at 300 K by He et al. \cite{He12} using EMD are in agreement with our values. Therefore, it is verified that although in AEMD one uses typically elongated supercells, with a very extreme aspect ratio between the $L_Z$ and $L_X,L_Y$ length (up to 900), a true 3D bulk conductivity is extracted, and not a 1D-reduced value, which would lead necessarily to an underestimation of $\kappa$. Finally, it can be seen that the results obtained by lattice dynamics (LD) by computing the three-phonon scattering at 300 K lie in the upper interval of the results found by AEMD. This is due to the fact that MD simulations implicitly account for phonon-phonon scattering mechanisms at any order, and not only for three-phonon collisions as it was the case in Ref. \onlinecite{He12}.

\begin{figure}[htb]
\includegraphics[width=\linewidth]{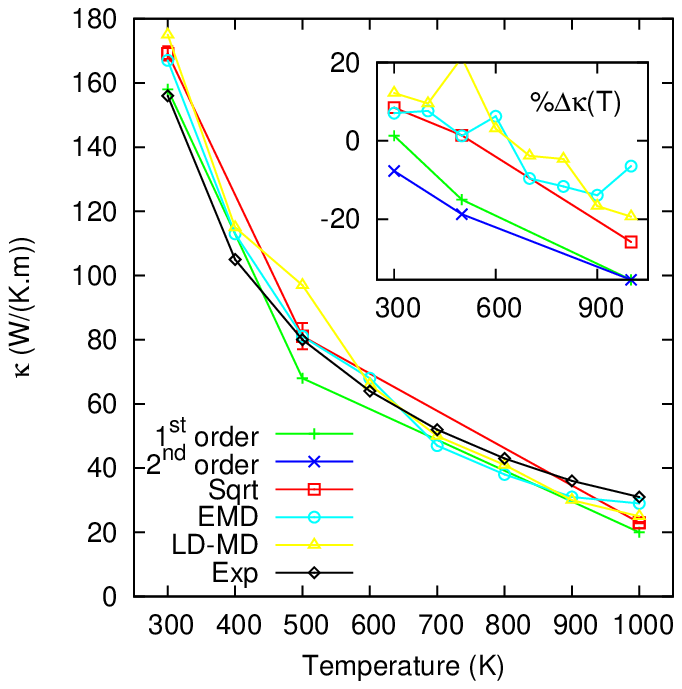}
\caption{Si bulk thermal conductivity calculated using EDIP  potential and various molecular dynamics approaches: AEMD extrapolated  at 1$^{\mathrm{st}}$ and 2$^{\mathrm{nd}}$ order formulations, and by the \emph{Sqrt}. EMD and mixed LD and MD calculations (``LD-MD'') \cite{Henry08}. In inset, $\kappa_\mathrm{theo}-\kappa_\mathrm{exp}$ in percentage of $\kappa_\mathrm{exp}$.} 
\label{FigkinfEDIP}
\end{figure}

The agreement between the calculation of the thermal conductivity by the various methods becomes even better with the EDIP potential, as shown in Fig. \ref{FigkinfEDIP}. The same conclusions already drawn using the Tersoff potential, i. e.  a slight difference between the bulk thermal conductivities obtained from Eq. \ref{ExtraSchell} or \ref{EqFit}, and good agreement with the EMD results are again obtained. In this case the agreement with combined LD-MD calculations by Henry and Chen\cite{Henry08} is even better. Moreover, all the values at different temperatures appear to be in close agreement with the experiment, as it can be seen in the inset, with an offset smaller than 10 \% at 300 K and 30\% at 1000 K. The larger offset at higher temperature is due to atomic vibrations sampling the energy landscape further away from the minimum, in an increasingly anharmonic regime that is difficult to account by using empirical interatomic potentials.

\begin{figure}[h]
\includegraphics[width=\linewidth]{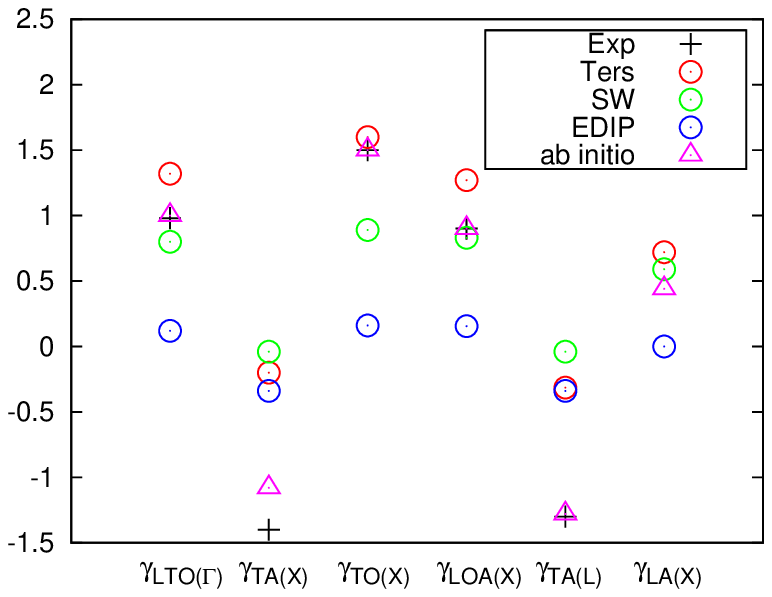}
\caption{Gr\"uneisen parameters of silicon, measurements (``Exp'') and calculations using Tersoff (``Ters''), Stillinger-Weber (``SW'') and EDIP (``EDIP'') potentials, and ab initio (``ab initio''). Data from Ref. \onlinecite{Porter97}.} 
\label{FigGrun}
\end{figure}

The apparently better agreement with experimental values obtained by using the EDIP potential is not easy to justify. Neither the phonon spectrum, and consequently the phonon velocity distribution, nor the heat capacity calculated by Porter et al. \cite{Porter97}, display any quantitative differences, whenever the Tersoff, Stillinger-Weber and EDIP potentials are used. Concerning the relaxation times, or mean free paths, these are intimately related to the ability of the interatomic potential to predict the anharmonicity, which is quantified by the Gr\"uneisen parameters. The same authors show that none of the three interatomic potentials are able to predict the experimental values of the 
Gr\"uneisen parameters (Fig. \ref{FigGrun}); in particular, for the acoustic modes these coefficients are not enough negative to match the experiments. Care must therefore be taken, in that this agreement of the thermal conductivity obtained using EDIP with the experimental values is likely to be fortuitous. 

\section{Conclusion}

In this work, we carried out an extensive study of the length dependence of the thermal conductivity in atomistic molecular dynamics simulations. Such a length dependence is typically observed in simulations in which boundary scattering is present by construction of a supercell, as well as in simulations methods (such as our AEMD) in which no sources of boundary scattering apparently exist. It is instead absent in other techniques, such as equilibrium MD and lattice dynamics that, although more expensive computationally and limited to 3-dimensional homogeneous systems, can be used as benchmark for a comparison. 

The length dependence observed in the MD simulations is found to be more pronounced for materials with a higher thermal conductivity, or for the lower temperatures for any given material. By comparing the length dependence of the thermal conductivity, and the accumulation of the thermal conductivity with increasingly long phonon mean free paths, we obtain a good correlation between the two. This demonstrates a physical correspondence between the kind of numerical cut-off imposed by the supercell finite length, and the maximum phonon mean free paths sampled in the simulation of heat diffusion. 

In a second part of our study, we used the idea that the length of the AEMD supercell acts as a cut-off for the maximum phonon mean free path, to derive a physically-motivated extrapolation scheme. A semi-analytical model, which assumes a length dependence equal to the MFP accumulation, and phonon MFP limited only by the intrinsic scattering by Umklapp, is shown to capture well the long-wavelength length-dependence in AEMD, over a large interval of temperatures, and for materials ranging from bad to good heat conductors. 

This analytical model was further used to extrapolate the thermal conductivity of silicon at infinite length, since in this material the phonon MFPs are too large even for the biggest MD simulations. The new formulation avoids the misuse of previously established extrapolation schemes, which could only be justified in particular MD simulations such as the ``direct method". We show that, for a given interatomic potential, the extrapolated values are in agreement with other extrapolations obtained from the ``direct method", and with simulation methods free of any length dependence. The additional conclusion is therefore that the strongly elongated supercell used in the AEMD method do not prevent to extract a correct bulk value of the thermal conductivity. 

Finally, we also showed that AEMD simulations using the EDIP potential could provide values of $\kappa$ for silicon that are the closest to the experiment, besides such an agreement could not be justified by the anharmonicity of this potential, as quantified by the Gr\"uneisen parameters. 

\section{Acknowledgements}
This work was funded by the French ANR via the project n. ANR-13-NANO-0009 ``NOODLES" and by Equipex EXCELSIOR (http://www.excelsior-ncc.eu). The computations were performed by using resources from GENCI (Grand Equipement National de Calcul Intensif) (Grant No. [2015]096939, [2014]096939, [2013]099029 and 100356). 


\end{document}